\documentclass[twocolumn,showpacs,showkeys,prb]{revtex4}
\usepackage{graphicx}
\begin{document}
\title{Theory of Doping Induced High-Spin in a Model of Polyene-Based Molecular Magnets}
\author{Ping Huai}
\author{Yukihiro Shimoi}
\author{Shuji Abe}

\affiliation{Nanotechnology Research Institute (NRI)  and  
Synthetic Nano-Function Materials Project (SYNAF),  \\ National Institute of
Advanced Industrial Science and Technology (AIST), 1-1-1 Umezono, Tsukuba,
Ibaraki 305-8568}


\begin{abstract}
Control of intramolecular spin alignment  is studied theoretically in a model of polyene-based
molecular magnets in which delocalized $\pi$ electrons are 
coupled with localized radical spins.
In a  previous paper  [Phys. Rev. Lett. 90 (2003) 207203], we have demonstrated
that charge doping is an effective way to realize controllable high-spin 
in the $\pi$-conjugated  molecular magnets. 
In this paper, we clarify the dependence of spin-alignment on the 
exchange interaction between  the  localized spin and $\pi$ electron and the electron-electron interactions.
The antiferromagnetic exchange interaction plays a role different from  the
ferromagnetic counterpart in doped molecules.
To understand complex interplay of charge and spin degrees of freedom
in the doped systems, we carry out a systematic study 
on the phase diagram of spin alignment in the parameter space.
The mechanism of the spin alignment is discussed  based on the spin densities of $\pi$ electrons.
The calculated results are consistent with experiments, providing a theoretical basis for
the control of  spin alignment.
\end{abstract}

\maketitle

\section{Introduction}
\begin{figure}[tb]
\begin{center}
\includegraphics[width=7.0cm]{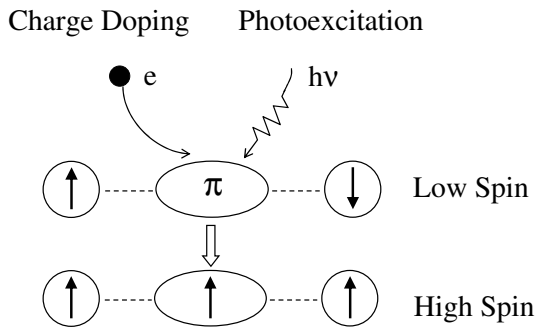}
\end{center}
\caption{Schematic picture of control of intramolecular spin alignment by charge doping or photoexcitation.}
\label{fig1}
\end{figure}

Since the discovery of high-spin ground state in dicarbene,\cite{itoh,wasserman}
molecular magnetism has been extensively investigated both experimentally and theoretically
for decades.\cite{rajca,miller,miller2,itoh2}
A new class of purely organic molecular magnets has received increasing attention 
as their intramolecular spin alignment is controllable by external stimuli such 
as charge doping or photoexcitation,\cite{izuoka1,sakurai,nakazaki,teki1,teki2,teki3,teki4,matsuda1,matsuda2,matsushita} 
as schematically shown in Fig. \ref{fig1}. 
Those molecules share a common feature in structure: two stable radicals carrying unpaired electrons
coupled with a $\pi$-conjugated moiety through exchange interaction. 
The two radicals are barely coupled to each other directly, but 
interact indirectly via $\pi$ electrons.
When the electronic state of 
the $\pi$-conjugated moiety is modulated by external stimuli,
these radicals change from
antiferromagnetic to ferromagnetic alignment, leading to a high-spin state.

Among the pioneering works, Izuoka \textit{et al.}  have successfully shown  that intramolecular spin
alignment  can be controlled by charge doping in 
thianthrene bis(nitronyl nitroxide).\cite{izuoka1}
The molecule is spin singlet ($S=0$) in neutral state with very weak antiferromagnetic exchange coupling 
($-1.4$ K) between the nitronyl nitroxide groups.
Upon one-electron oxidation, \textit{i.e.} one-hole doping,
spin quartet ($S=3/2$) ESR signals are detected in the temperature range of $5-150$ K, 
indicating a low-spin to high-spin transition induced by doping.
Optically controllable spin alignment has been demonstrated by
Teki \textit{et al.} in a molecule composed of dangling iminonitroxide radicals 
and $\pi$-conjugated moiety of diphenylanthracene.\cite{teki1,teki2} 
The ground state of the molecule is determined to be a low-spin
state, corresponding to weak antiferromagnetic coupling ($-5.8$ K) between the radical spins.
After the molecule is excited in low temperature ($20-40$ K) by short pulses of laser,
spin  quintet ($S=2$) species are observed in  $10~\mu$s by using
the time-resolved ESR spectroscopy. This metastable spin  quintet results from
photoinduced parallel spin alignment among the two stable radicals and a photogenerated triplet 
$\pi$ radical.
Since these molecular magnets are purely organic, there is 
a great advantage of various chemical modifications onto the $\pi$-conjugated moiety.
It might allow such molecular magnets 
to be building blocks of photoinduced bulk magnets similar to  Prussian-blue analogs.\cite{sato,kawamoto}

Many theoretical works have been carried out to propose and understand the mechanism
of spin alignment in molecular magnets
\cite{mataga,ovchinnikov,yamanaka,mizouchi,tyutyulkov,fukutome,sinha,yoshizawa,arita,nasu,fang,chen}.
By a weak electron-correlation approach, Mataga predicted that ferromagnetic spin alignment
appears in several hypothetical hydrocarbons with
$\pi$-conjugated electron systems due to the topological nature of molecular orbitals.\cite{mataga}
Ovchinnikov accounted for the net spins of planer alternate hydrocarbons 
by a strong electron-correlation approach.\cite{ovchinnikov}
These two approaches were later generalized to rigorous theorems in the Hubbard model 
by Lieb.\cite{lieb}
Thus the dominant mechanism of intramolecular spin alignment in the ground state
of $\pi$-conjugated  molecules has been well established  by those works:
The alignment is governed by the topological rule
based on the dynamical antiferromagnetic spin polarization effect of $\pi$ electrons
with on-site Coulomb repulsion. 

The topological rule can be applied only to the half-filled  ground states, but 
spin alignment in doped or excited molecules remains unsolved for general cases. 
In the strong-correlation limit with an infinite value of on-site Coulomb
repulsion in the Hubbard model, Nagaoka proved rigorously that
the ferromagnetic ordering is favorable 
in the case of  single-electron addition to or removal from half-filling.\cite{nagaoka}
Most discussions on the spin-alignment control so far have
been given on the basis of individual molecular orbitals.\cite{sakurai,teki2,teki4}
For high-spin states induced by charge doping, Sakurai \textit{et al.}
interpreted parallel alignment of two radical spins in terms of the electronic structures
calculated by the unrestricted Hartree-Fock theory.\cite{sakurai}
A mechanism of photoexcited high-spin states was discussed by Teki \textit{et al.} using 
ab initio molecular orbital calculations based on 
the density functional theory.\cite{teki2,teki4} 

Approaches by model Hamiltonians have been widely used to study the  neutral ground state
of quasi-one-dimensional (1d) organic ferromagnets. 
Nasu \cite{nasu} studied spin alignment in an idealized poly(m-diphenylcarbene) 
using a periodic Kondo-Hubbard model. 
Fang \textit{et al.}\cite{fang} and Chen \textit{et al.}\cite{chen} studied a periodic 
Kondo Su-Schrieffer-Hegger(SSH) model\cite{ssh} taking into account the strong
electron-phonon interaction as well as the electron-electron correlation. 
In the previous papers \cite{huai,huai2}, we introduced a
Kondo-Peierls-Hubbard model to study spin alignment of $\pi$-conjugated molecular magnets. 
It has been demonstrated that a spin singlet to quartet transition can be induced by electronic doping into
the $\pi$ moiety, which is ferromagnetically coupled with localized spins.
The intramolecular spin alignment depends sensitively on the topological structure
of the molecular system. The doped high-spin state can be realized
in appropriate molecular structures.
The topological effect in the doped case is very different from that in the half-filled case.

The aim of the present paper is to clarify the dependence of spin-alignment
on spin exchange and electron-electron interactions, thereby giving an integral picture of
spin-alignment control in polyene-based  molecular magnets.
The paper is organized as follows:
In Sec. \ref{model}, we introduce  the model and the calculation method.
The calculated results are discussed in Sec. \ref{results}.
We first study the spin alignment and the lattice deformation pattern with
antiferromagnetic coupling as well as the ferromagnetic one in Sec. \ref{sub1}.
The dependence of spin alignment on the molecular structure
is discussed in Sec. \ref{sub2}.
To study the complex interaction between charge and spin
in doped quantum systems, we discuss the dependence of spin alignment on parameters $J$ and $U$ 
in Sec. \ref{sub3}. All the results are summarized in Sec. \ref{summary}. 

\section{Kondo-Peierls-Hubbard Model}
\label{model}
In this paper, we use the same Kondo-Peierls-Hubbard model as in our previous papers \cite{huai,huai2}:
a Peierls-Hubbard model on a $N$-site linear chain 
interacting with two localized quantum spins in the form of the Kondo coupling.
The open-boundary linear chain corresponds 
to polyene, and its $\pi$ electrons are subject to electron-lattice interactions in the form of the SSH coupling.\cite{ssh} 
The localized spins correspond to the unpaired electrons of stable radical groups 
attached to the main chain, as realized in some substituted polyacetylenes.\cite{nishide,iwamura}
The polyene-based molecular magnets described by this model is currently hypothetical
but there is a possibility to synthesize them in  future.

Thus, our model is presented in the Hamiltonian,
\begin{eqnarray}
H &=& -\sum_{i,s} t_i (C^{\dagger}_{i,s}C_{i+1,s} + h.c. )  \nonumber \\
  & & + U\sum_{i} n_{i,\uparrow} n_{i,\downarrow} 
    + \frac{K}{2}\sum_{i} (q_i -q_{i+1})^2  \nonumber \\
  & &  - J (\mathbf{S}_{i1} \cdot  \mathbf{S}_{\text{T1}}+
           \mathbf{S}_{i2} \cdot  \mathbf{S}_{\text{T2}})~,   \\
t_i &\equiv& t_0 + \alpha(q_i-q_{i+1}) ,\\ 
\mathbf{S}_{i}&=&\frac{1}{2} C^{\dagger}_{i} \mathbf{\sigma} C_{i}.
\end{eqnarray}
In Eq. (1),  $C^{\dagger}_{i,s}$($C_{i,s}$) creates (annihilates) an electron with spin 
$s$($=\uparrow$ or $\downarrow$) on site $i$. The  transfer integral $t_i$ is 
introduced for the nearest-neighbor sites, and is
composed of a constant magnitude $t_0$, and a variable one that depends
on the displacement $q_i$ of site $i$ with the SSH coupling constant $\alpha$. 
On-site Coulomb repulsion energy is given by $U$, and $n_{i,s}\equiv C^{\dagger}_{i,s}C_{i,s}$. 
The elastic constant of the lattice is $K$, while the kinetic energy of lattice is neglected.
The last term of Eq. (1) is the exchange coupling of two
localized $1/2$ spins ($\mathbf{S}_{\text{T1}}$ and
$\mathbf{S}_{\text{T2}}$) to the spins of $\pi$ electrons at site $i1$ and $i2$, respectively.
$C^{\dagger}_{i}$ is defined as $(C^{\dagger}_{i,\uparrow},C^{\dagger}_{i,\downarrow})$ and 
$\mathbf{\sigma}\equiv (\sigma_x,\sigma_y,\sigma_z)$ are the Pauli matrices.
We define a dimensionless coupling constant $\lambda \equiv \alpha^2/(t_0 K)$ 
and dimensionless bond-lengths $\Delta_i \equiv  \alpha(q_i- q_{i+1})/t_0$.

We exactly diagonalize the electronic and spin part of the Hamiltonian by the Lanczos algorithm. 
Therefore the obtained wavefunctions take account of all the correlation effects.
The lattice deformation is treated classically and optimized by means of 
the Hellmann-Feynman force equilibrium condition at zero temperature,
\begin{eqnarray}
\Delta_i = \lambda \left <  \sum_{s} (C^{\dagger}_{i,s}C_{i+1,s} + h.c. ) \right > ,
\end{eqnarray}
where $\langle \cdots \rangle$ denotes the ground state expectation value. 
We perform these two procedures iteratively until the electronic state and lattice deformation converge.
Since the model has the electron-hole symmetry, electron doping gives the same results as
hole doping. Only the results for hole doping will be discussed in the  paper.

First let us analyze the  case of strong-correlation limit $U/ t_0 \rightarrow \infty $.
The half-filled Hubbard model is mapped onto the Heisenberg model with 
effective exchange interactions $J^{\rm eff}_i = -4t_i^2/U$ ($<0$),
so that the electronic and spin parts of 
the present Hamiltonian is reduced to 
\begin{eqnarray}
H_{eff} &=&- \sum_{i} J^{\rm eff}_i (\mathbf{S}_i \mathbf{S}_{i+1} - 1/4) 
        \nonumber \\ 
    & &- J (\mathbf{S}_{i1} \cdot \mathbf{S}_{\text{T1}} + \mathbf{S}_{i2} \cdot \mathbf{S}_{\text{T2}}) .
\end{eqnarray}
The spin alignment predicted by the topological rule can be deduced from
such a spin Hamiltonian in an intuitive way,
if quantum fluctuations are neglected. 
Because of antiferromagnetic correlation between the neighboring $\pi$ spins, 
the alignment between $\mathbf{S}_{T1}$ and $\mathbf{S}_{T2}$ 
can be classified in terms of the parity of $i2-i1$:
antiparallel if $i2-i1$ is odd, but parallel if $i2-i1$ is even,
no matter whether the exchange coupling $J$
is ferromagnetic or antiferromagnetic. For $J<0$, this intuitive picture
is consistent with rigorous theorems.\cite{lieb-mattis,ovchinnikov}
On the other hand, single hole doping induces Nagaoka ferromagnetic ordering 
in the limit of strong correlation.\cite{nagaoka} Therefore the two localized spins
favor parallel alignment.
If $U/t$ is finite, no simple rule exists for the doped  $\pi$ electron systems,
and numerical methods are more effective way to study doped states.

The model is investigated with the following parameters in the unit of $t_0$ (on the order of eV):
$U=2.5$, $|J|=0.2$ while $\lambda$ is allowed to vary from 0 to 0.5. 
In the previous paper \cite{huai}, we assumed the exchange interaction much stronger ($|J|=1$)
than this value, and discussed mostly the case of ferromagnetic coupling.
Here, weaker exchange interaction is applied to the model because the reduced value gives
more realistic spin gap.
Furthermore, $J$ will be discussed for both ferromagnetic (F) $J>0$ and antiferromagnetic (AF) $J<0$,
although $J$ has been suggested to be positive in the thianthrene-based molecular magnets \cite{sakurai}.

\section{Results and Discussion\label{results}}
\subsection{Spin correlations and lattice deformations\label{sub1}}
\begin{figure}[tb]
\begin{center}
\includegraphics[width=7.0cm]{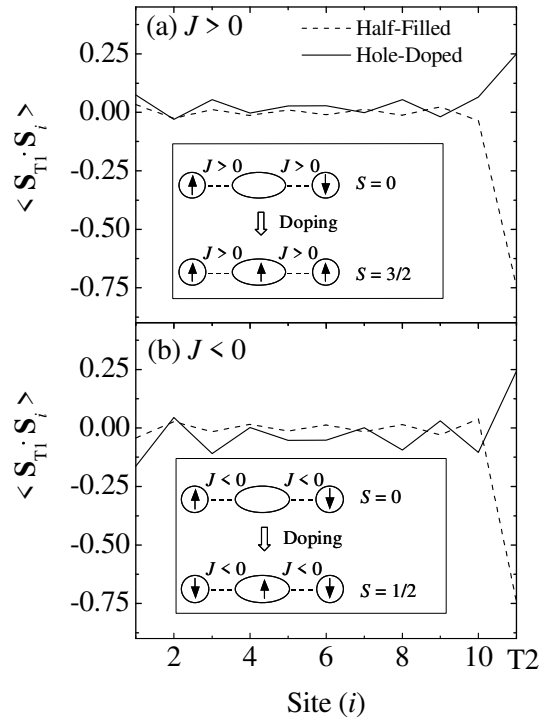}
\end{center}
\caption{Spin correlations $\langle \mathbf{S}_{\text{T1}} \cdot \mathbf{S}_i \rangle$ for 
the half-filled ($N_e=N$; dashed line) and the single hole doped cases ($N_e=N-1$; solid line)
with $|J|=0.2$, $U=2.5$, $\lambda=0.25$, and $N=10$.
(a) F coupling ($J>0$) and (b) AF coupling ($J<0$).
The inset in each figure shows schematically the change of spin alignment by hole doping.}
\label{fig2}
\end{figure}

To investigate  spin alignment with electron-lattice interaction,
we first study a $10$-site chain with $\lambda=0.25$ (a typical value for polyacetylene 
\cite{ssh,jeckelmann}).
The two localized spin are coupled to electron spin at both ends of the chain, 
 \textit{i.e.} $i1=1$ and $i2=10$.
The spin correlations are plotted in Fig. \ref{fig2} for both the half-filled ($N_e=10$)
and the single hole doped ($N_e =9$) cases.  

The correlations between $\mathbf{S}_{\text{T1}}$ and the other spins are shown in 
Fig. \ref{fig2}(a) for the case of F coupling ($J>0$). 
The correlation between $\mathbf{S}_{\text{T1}}$
and $\mathbf{S}_{\text{T2}}$ is  antiferromagnetic for the half-filled case,
resulting in spin singlet $S=0$. 
This antiparallel spin alignment can be understood based on 
the topological rule for the half-filled ground state. The spin correlation
($\langle \mathbf{S}_{\text{T1}} \cdot \mathbf{S}_{\text{T2}} \rangle=-0.745$) 
is near to $-3/4$ of a singlet pair of the localized spins.

Positive correlation appears between the localized spins when the $\pi$ electron moiety is doped by a hole.
Similar spin alignment has been reported in the case of the stronger coupling \cite{huai}. 
In the present case, this spin
correlation ($\langle \mathbf{S}_{\text{T1}} \cdot \mathbf{S}_{\text{T2}} \rangle=0.249$)
is  near to $1/4$ of a triplet pair of the localized spins.
The spin correlation pattern in the right half of the chain ($i \geq 6$) is also reversed by
the doping.
Furthermore, the doped system turns out to be a spin quartet ($S=3/2$). 
The inset in Fig. \ref{fig2}(a) illustrates the alteration of spin alignment from the spin singlet to
the spin quartet by such hole doping. 
This kind of low-spin to high-spin transition corresponds to 
the observation in the aforementioned thianthrene derivative.\cite{izuoka1}

Figure \ref{fig2}(b) shows the spin correlation in the case of AF coupling ($J<0$).
Antiferromagnetic correlation between  $\mathbf{S}_{\text{T1}}$ and
$\mathbf{S}_{\text{T2}}$ ($=-0.742$) is found in the half-filled case with the total spin $S=0$, which
is also expected from the topological rule. However,
hole doping induces a spin doublet $S=1/2$, while the correlation 
($\langle \mathbf{S}_{\text{T1}} \cdot \mathbf{S}_{\text{T2}} \rangle=0.243$) is ferromagnetic similar
to the case of F coupling.
The antiparallel alignment between the hole spin and 
the localized spins is responsible for the spin doublet
as schematically shown in the inset of Fig. \ref{fig2}(b). 

\begin{figure}[tb]
\begin{center}
\includegraphics[width=7.0cm]{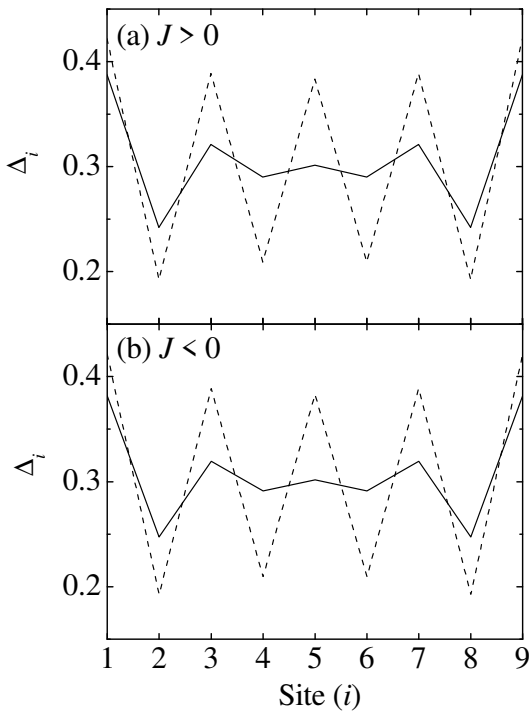}
\end{center}
\caption{Lattice deformations $\Delta_i$ for the half-filled ($N_e=N$; dashed line) 
and the single hole doped cases ($N_e=N-1$; solid line) 
with $|J|=0.2$, $U=2.5$, $\lambda=0.25$, and $N=10$.
(a) F coupling ($J>0$) and (b) AF coupling ($J<0)$.}
\label{fig3}
\end{figure}

The lattice deformation patterns is plotted in Fig. \ref{fig3} for both F and AF couplings.
Quite similar profiles are obtained in both cases: half-filled case
has bond alternation in the lattice, while the doped case contains
very weak bond alternation especially around the middle of the lattice. 
The localization of the lattice relaxation is ascribed to a polaronic effect as well
as a  chain-end effect as shown in ref. \onlinecite{kuwabara}.

\begin{figure}[tb]
\begin{center}
\includegraphics[width=7.0cm]{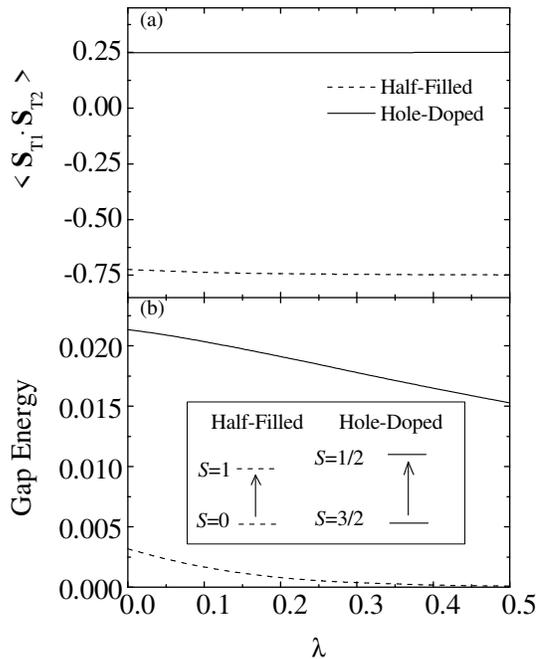}
\end{center}          
\caption{(a) Spin correlations $\langle \mathbf{S}_{\text{T1}} \cdot \mathbf{S}_{\text{T2}} \rangle $ and
         (b) gap energies as functions of $\lambda$ for the half-filled ($N_e=N$; dashed line)
          and the doped ($N_e=N-1$; solid line) cases with 
          $J=0.2$, $U=2.5$, $\lambda=0.25$, and $N=10$.
          The inset in (b): the total spins of the ground and the first excited states.        
          }
\label{fig4}
\end{figure}

We examine the effect of electron-lattice coupling on spin alignment by investigating  
spin correlation and gap energy between the ground state and the first excited state
in the range of $0 \leq \lambda \leq 0.5$. The $\lambda$-dependences
are plotted  in Fig. \ref{fig4} for F coupling. 
For both the half-filled and the doped cases, the spin correlation is barely affected by $\lambda$.
The inset in Fig. \ref{fig4}(b) indicates that
the lowest excitation takes place from singlet to triplet in the half-filled 
case, and from quartet to doublet in the doped case.
In the half-filled case with $\lambda=0.25$, 
the spin gap   ($5.5 \times 10^{-4}~t_0$, about a few  K) 
is comparable to the effective couplings of thianthrene bis(nitronyl nitroxide) \cite{izuoka1}.
The spin gap gradually decreases as the electron-lattice coupling increases, 
because of the increasing tendency of local singlet-pair formation in $\pi$ electrons.
The gap in the doped case is an order of magnitude larger than that 
in the half-filled case, and decreases as $\lambda$ increases.
Spin polarization is suppressed by strong bond alteration
as electron-lattice coupling increases.
The spin gaps of very weakly coupled systems tend to decrease 
with reduction of spin densities on the coupling sites.  
This $\lambda$-dependence of the spin gap is different from that reported
in our previous paper with a larger $J$ \cite{huai}.
In the case of strong exchange interaction, the localized spins are firmly bound to
the $\pi$-spins, and are hardly set free by low-energy excitations. 
Thus the spin gap would be mainly governed by excitations within the 
$\pi$-moiety. These $\pi$-excitation energies are raised by the electron-lattice coupling,
resulting in the increase  of the  spin gap with $\lambda$ in the stronger $J$ case.

\begin{figure}[tb]
\begin{center}
\includegraphics[width=7.0cm]{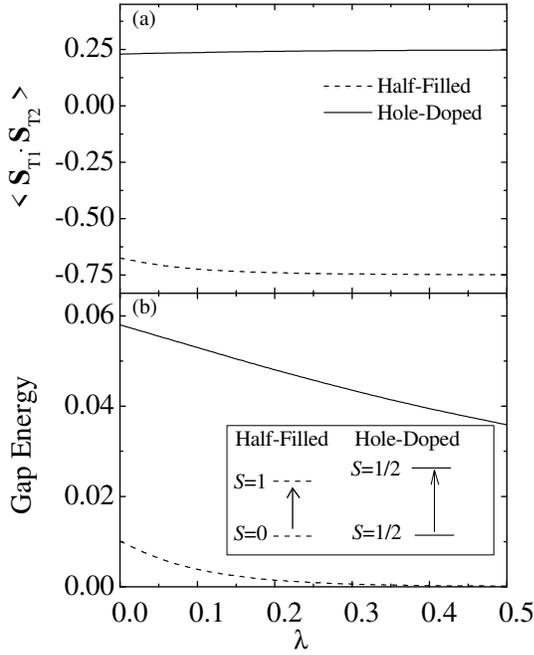}
\end{center}
\caption{(a) Spin correlations $\langle \mathbf{S}_{\text{T1}} 
         \cdot \mathbf{S}_{\text{T2}} \rangle$ and
         (b) gap energies as functions of $\lambda$ for the half-filled 
         ($N_e=N$; dashed line)
          and the doped ($N_e=N-1$; solid line) cases with 
          $J=-0.2$, $U=2.5$, $\lambda=0.25$, and $N=10$.
          The inset in (b): the total spins of the ground and the first excited states.
          }
\label{fig5}
\end{figure}

For the AF coupling, the spin correlation and the gap energy are plotted 
as functions of  $\lambda$ in Fig. \ref{fig5}.
We see tendencies similar to the case of the  F coupling,
even though the manner of spin alignment in the doped state
depends on the sign of the coupling,
as already seen in the insets of  Fig. \ref{fig2}.
In the half-filled case with $\lambda=0.25$, 
the spin gap ($9.5 \times 10^{-4}~t_0$) is larger than that of the F coupling but is still 
fairly small against thermal fluctuations at room temperature.
The suppression of electron-lattice coupling is desirable in molecular design with respect to
the thermal stability.

\subsection{Effects of molecular structure}
\label{sub2}
Geometrical structures are very important to $\pi$-conjugated 
molecular magnets. The influence on the doped state is  different from that on the neutral one,
as pointed out in the previous paper \cite{huai}. 
It has been observed in the topological isomers of thianthrene bis(nitronyl nitroxide),
that spin state depends on the positions of the radicals in the neutral case 
but has no position-dependence in the doped case \cite{izuoka1}.
In this paper, we examine two geometrical factors:
the length of the chain $N$ and the positions of the localized spins. 

\begin{figure}[tb]
\begin{center}
\includegraphics[width=7.0cm]{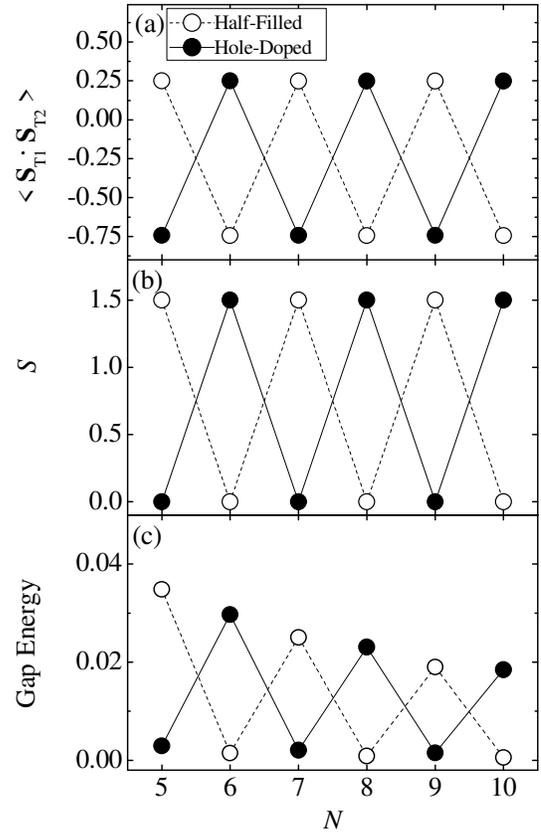}
\end{center}        
\caption{(a) Spin correlations $\langle \mathbf{S}_{\text{T1}} \cdot 
          \mathbf{S}_{\text{T2}} \rangle$, 
         (b) total spins, and 
         (c) gap energies as  functions of $N$ for the half-filled ($N_e=N$; open circle)
          and the doped ($N_e=N-1$; filled circle) cases with 
          $J=0.2$, $U=2.5$, and $\lambda=0.25$.
        }
\label{fig6}
\end{figure}

First we examine  the $N$-dependence of  spin alignment. 
Figure \ref{fig6} shows the correlations 
$\langle \mathbf{S}_{\text{T1}} \cdot \mathbf{S}_{\text{T2}}\rangle$ 
between the localized spins, the total spins, 
and the gap energies as functions of $N$ in the case of the F coupling.
To focus on the effect of the length, we maintain the localized spins at both ends of the chain.
In the half-filled case, the topological rule governs the spin correlation and total spin  perfectly.
For even number of $N$, spin state is singlet ($S=0$)
with antiparallel alignment between $\mathbf{S}_{\text{T1}}$ and $\mathbf{S}_{\text{T2}}$,
while for the odd number case total spin is quartet ($S=3/2$) with parallel alignment. 

Similar alternating behavior is also seen in the hole-doped systems, 
although the topological rule is generally only applicable for the ground state 
of the half-filled system.
The hole doping changes the total spin and spin correlation 
$\langle \mathbf{S}_{\text{T1}} \cdot \mathbf{S}_{\text{T2}}\rangle$ as follows:
$S=0 \rightarrow S=3/2 $ with 
$\langle \mathbf{S}_{\text{T1}} \cdot \mathbf{S}_{\text{T2}}\rangle >0 \rightarrow
 \langle \mathbf{S}_{\text{T1}} \cdot \mathbf{S}_{\text{T2}}\rangle <0$, if $N$ is even; 
$S=3/2 \rightarrow S=0$ with 
$\langle \mathbf{S}_{\text{T1}} \cdot \mathbf{S}_{\text{T2}}\rangle <0 \rightarrow
 \langle \mathbf{S}_{\text{T1}} \cdot \mathbf{S}_{\text{T2}}\rangle >0$, if $N$ is odd.
It is noted that the spin correlation does not decay with increasing
the chain length both in the doped and half-filled cases, 
while the spin gap decreases gradually as shown in Fig. \ref{fig6}(c).
In general, one-dimensional electron systems have no long-range charge (spin)
order as a result of quantum fluctuations.
The correlation between a pair of electrons decays in a manner of power law with increasing distance.
Decreasing electron correlation would weaken the interaction between the localized spins
in our model, and reduce the spin gap as the chain becomes longer.

\begin{figure}[tb]
\begin{center}
\includegraphics[width=7.0cm]{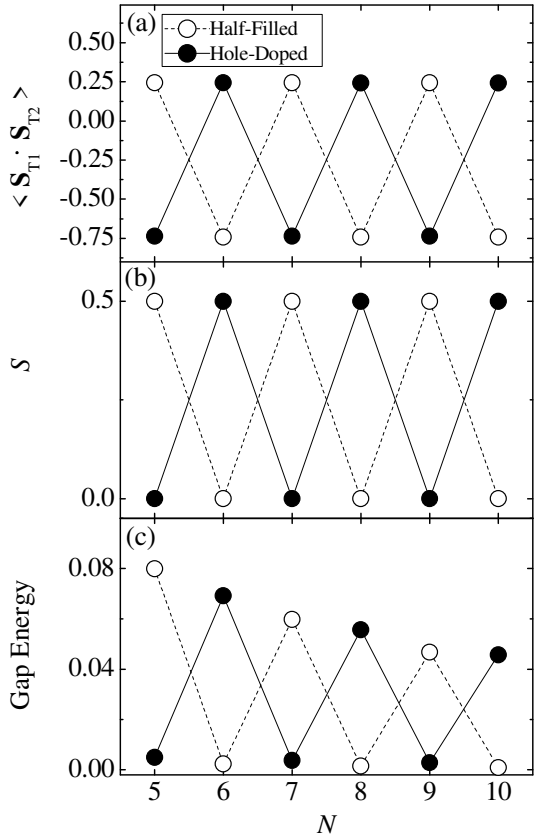}
\end{center}        
\caption{(a) Spin correlations $\langle \mathbf{S}_{\text{T1}} \cdot 
          \mathbf{S}_{\text{T2}} \rangle $, 
         (b) total spins, and 
         (c) gap energies as  functions of $N$ for the half-filled ($N_e=N$; open circle)
          and the doped ($N_e=N-1$; filled circle) cases with 
          $J=-0.2$, $U=2.5$, and $\lambda=0.25$.
        }
\label{fig7}
\end{figure}

We turn to the case of the AF coupling shown in Fig. \ref{fig7}.
In the half-filled system, the alignment of $\mathbf{S}_{\text{T1}}$ and $\mathbf{S}_{\text{T2}}$
is determined by the parity of $N$ in the same manner of the F coupling:
antiparallel if $N$ is even,  parallel if $N$ is odd.
In contrast to the F coupling, the total spin is singlet ($S=0$)  for even $N$, and 
doublet ($S=1/2$) for odd $N$,  as shown in Fig. \ref{fig7}(b).
The single hole doping flips the localized spin from antiparallel to parallel
with $S=0 \rightarrow S=1/2 $ if $N$ is even,
and  from parallel to antiparallel with $S=1/2 \rightarrow S=0$ if $N$ is odd.
The $N$-dependence of the spin correlation and of the spin gap is similar to the case of the F coupling.

\begin{figure}[tb]
\begin{center}
\includegraphics[width=7.0cm]{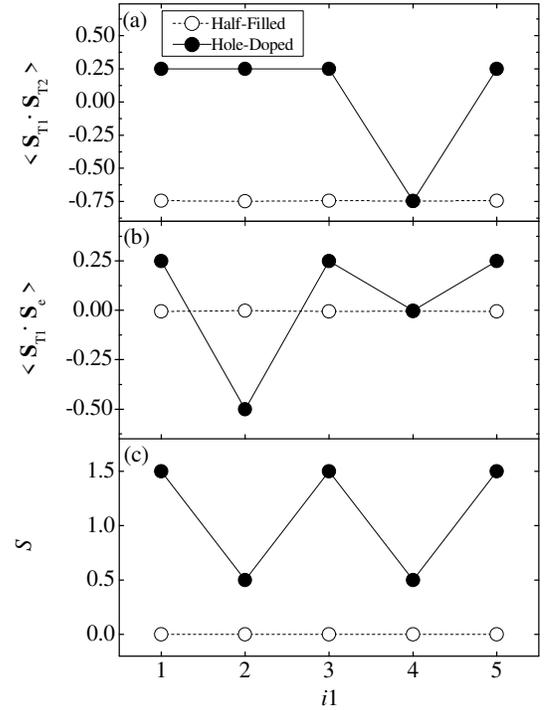}
\end{center}         
\caption{(a) Spin correlations $\langle \mathbf{S}_{\text{T1}} \cdot \mathbf{S}_{\text{T2}} \rangle $, 
         (b) $ \langle \mathbf{S}_{\text{T1}} \cdot \mathbf{S}_{\text{e}} \rangle $, and
         (c) total spins as functions of $i1$, with $i2=N+1-i1$, 
         for the half-filled ($N_e=N$; open circle) and the doped 
         ($N_e=N-1$; filled circle) cases with
         $J=0.2$, $U=2.5$, $\lambda=0.25$, and $N=10$. 
         }
\label{fig8}
\end{figure}

Next we proceed to the other geometrical factor, considering
the localized spins attached to inner sites of
the chain. We pick up symmetric positions of the localized spins as
$(i1,i2)=(1,N), (2,N-1), \cdots$, with $N=10$. 
In Fig. \ref{fig8}, the spin correlations $\langle \mathbf{S}_{\text{T1}} \cdot \mathbf{S}_{\text{T2}} \rangle $
, $\langle \mathbf{S}_{\text{T1}} \cdot \mathbf{S}_{\text{e}} \rangle$ 
($\mathbf{S}_{\text{e}} \equiv  \sum_{i=1}^N {\mathbf{S}_i}$),
and the total spin are plotted as functions of $i1$ for the F coupling. 
In the half-filled case, 
these values are almost independent of $i1$:
the spin alignment is antiparallel
with $S=0$, as expected by the topological rule, since 
$i2-i1$ is odd. The correlation between $\mathbf{S}_{\text{T1}}$ and  $\mathbf{S}_{\text{e}}$ is
negligible, because $\langle \mathbf{S}_{\text{e}} \rangle $ is almost zero
in the half-filled electron systems.

For the doped case,
the most interesting feature  is that the relative alignment among the
two localized spins and the hole spin depends on $i1$ and $i2$.
In Fig. \ref{fig8}(a), the localized spins 
exhibit positive correlation $\langle \mathbf{S}_{\text{T1}} \cdot \mathbf{S}_{\text{T2}} \rangle$
except for the position $(i1,i2)=(4,7)$.
Furthermore, these positively correlated spins can be classified into two 
categories in terms of $\langle \mathbf{S}_{\text{T1}} \cdot \mathbf{S}_{\text{e}} \rangle$, namely
the alignment between the localized spin and the  hole spin.
They are parallel in the case of $i1=1~,3,\text{~and}~5$
but antiparallel in the case of $i1=2$ (Fig. \ref{fig8}(b)).
Correspondingly the doped systems are spin quartet ($S=3/2$) 
for the former three structures but spin doublet ($S=1/2$) for the last one (Fig. \ref{fig8}(c)).
For $(i1,i2)=(4,7)$, the localized spins  are in antiparallel alignment leading to
spin doublet ($S=1/2$).

\begin{figure}[tb]
\begin{center}
\includegraphics[width=7.0cm]{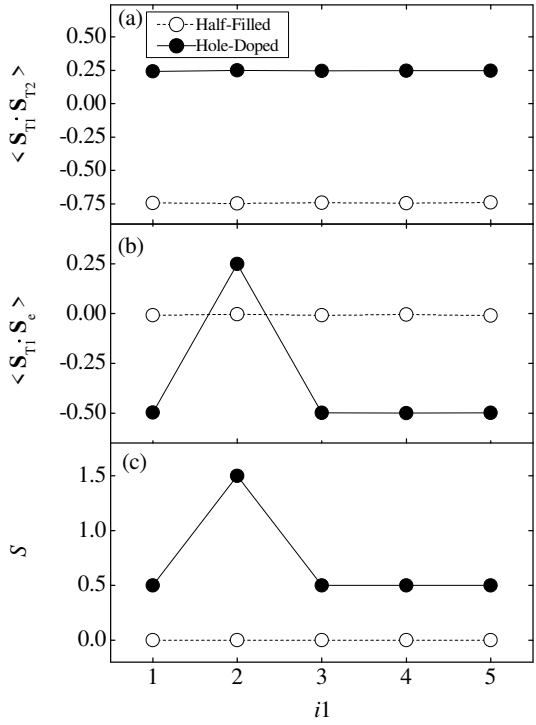}
\end{center}        
\caption{(a) Spin correlations $\langle \mathbf{S}_{\text{T1}} \cdot \mathbf{S}_{\text{T2}} \rangle $ and 
         (b) $\langle \mathbf{S}_{\text{T1}} \cdot \mathbf{S}_{\text{e}} \rangle$, and
         (c) total spins as functions of $i1$, with $i2=N+1-i1$, 
         for the half-filled ($N_e=N$; open circle) and the doped 
         ($N_e=N-1$; filled circle) cases with
         $J=-0.2$, $U=2.5$, $\lambda=0.25$, and $N=10$.
        }
\label{fig9}
\end{figure}

Figure \ref{fig9} shows results for the AF coupling as the localized spins are attached to inner sites of
the chain. Just like the F coupling,
the half-filled systems do not exhibit the $i1$-dependence: they are all spin singlet ($S=0$)
with antiparallel alignment between the
localized spins.
In the doped systems, the localized spins show parallel alignment
in all the cases as shown in Fig. \ref{fig9}(a).
On the other hand, they show antiparallel alignment with the hole spin
except for $ i1=2$ (Fig. \ref{fig9}(b)). 
This difference in spin alignment results in the different total spin: 
Spin doublet ($S=1/2$) 
for $i1=1~,3,4, \text{~and}~5$ and  spin quartet ($S=3/2$) for $i1=2$ (Fig. \ref{fig9}(c)). 

\begin{figure}[tb]
\begin{center}
\includegraphics[width=7.0cm]{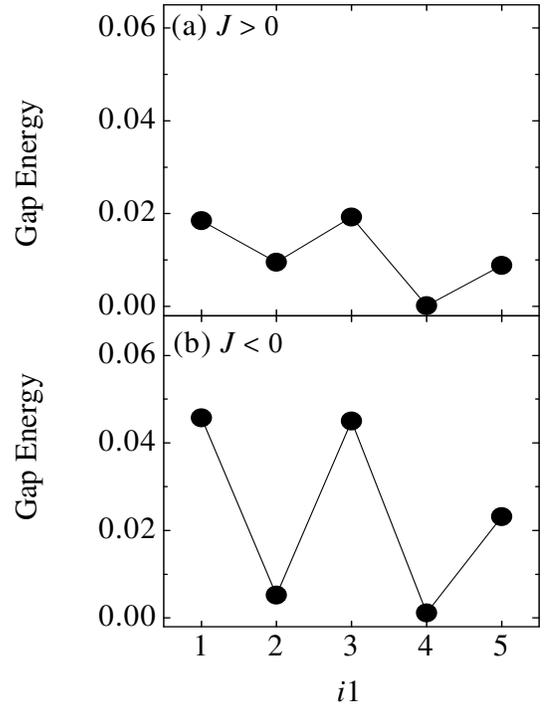}
\end{center}  
\caption{Gap energies of the doped system: (a) $J>0$ and (b) $J<0$
         as functions of $i1$ ($i2=N+1-i1$), with
         $|J|=0.2$, $U=2.5$, $\lambda=0.25$, and $N=10$.}
\label{fig10}
\end{figure}

The gap energies of the doped systems are presented in Fig. \ref{fig10}.
The gap is relatively large at $i1=1$ and 3  for both $J>0$ and $J<0$, 
where the high-spin states appear  in the case of $J>0$.
All the results above imply  that 
molecular systems with particular structures are suitable 
for the purpose of spin-alignment control since they exhibit doping-induced high-spin
and have large spin gaps.  

\subsection{Dependence on $J$ and $U$}
\label{sub3}
It is a key issue in the present study
to understand  the complex interplay of spin and
charge degrees of freedom in doped quantum systems. 
We have demonstrated the spin-alignment control by
hole-doping using fixed parameters of the on-site Coulomb repulsion $U$ and the exchange interaction $J$.
However, a subtle balance between these interactions governs the 
alignment  between the separated radical spins. 
It leads to a question: Does spin alignment change dramatically 
with varying the strengths of
$U$ and $J$, or just depend on them quantitatively?
As already shown in Fig. \ref{fig10},  the spin gap is sometimes very small in
the doped systems, hence 
the order of the present lowest excited and ground states could  be interchanged 
for another set of parameters. 
Furthermore, the spin alignment depends on the topological structure
as shown in Sec. \ref{sub2}. 
Thus, it is necessary to investigate the dependence on $U$ and $J$ 
for various molecular structures.
In the following discussion, the electron-lattice coupling is omitted 
as it plays less important role than the electron-electron interaction.

\begin{figure}[tb]
\begin{center}
\includegraphics[width=7.0cm]{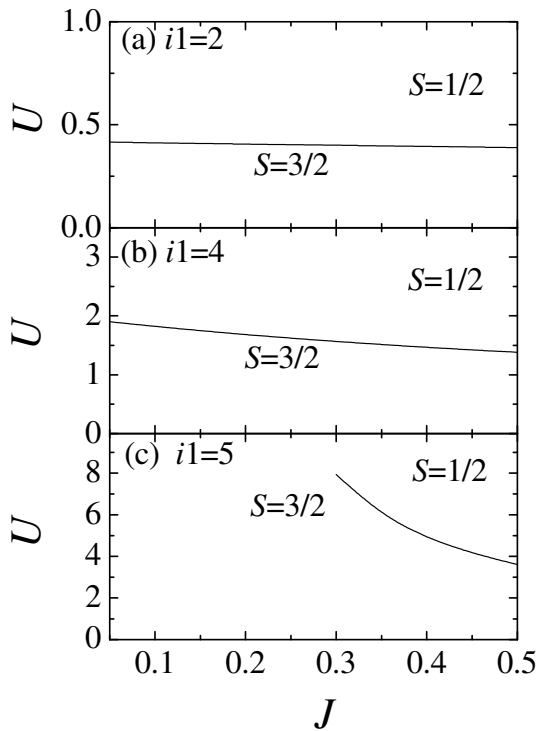}
\end{center}         
\caption{Phase diagrams of the doped system ($N_e=N-1$) for 
         (a) $i1=2$, (b) $i1=4$, and (c) $i1=5$, ($i2=N+1-i1$)
         with $J>0$, $\lambda=0$, and $N=10$. }
\label{fig11}
\end{figure}

In the case of F coupling, the phase diagrams in the parameter space of $J$ and $U$
are shown for $i1=2$, 4, and 5 in Fig. \ref{fig11}. 
The doped systems with $i1=1~\text{or}~3$ are always spin quartet 
in the  calculated range of $ 0 <J<0.5$ and $0 <U< 8$.
As the on-site Coulomb interaction increases over a critical value
about $U \sim 0.4$ for $i1=2$, or $U \sim 2.0$ for $i1=4$,
the systems are transferred from the spin quartet to the spin doublet,
and the phase boundaries are  almost independent of the exchange interaction.  
The system with $i1=5$  has much strong dependence on $J$ and $U$.
It also undergoes a transition
from the spin quartet to the spin doublet, but occurs in stronger Coulomb interaction $U>4.0$.
Detailed analysis reveals that the spin gap goes continuously to zero
when the system approaches to the phase boundary.  

\begin{figure}[tb]
\begin{center}
\includegraphics[width=7.0cm]{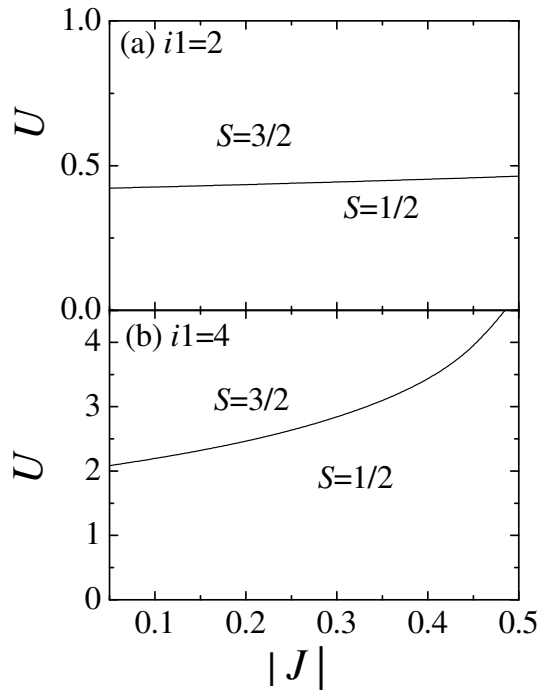}
\end{center}         
\caption{Phase diagrams of doped system ($N_e=N-1$) for (a)$i1=2$ and
          (b)$i1=4$,($i2=N+1-i1$)
         with $J<0$, $\lambda=0.0$ and $N=10$. }
\label{fig12}
\end{figure}

Similar transitions are  also found  in the case of AF coupling.
Figure \ref{fig12} shows the phase diagrams for the systems with $i1=2$ and $4$, which 
are transferred
from spin doublet to quartet at $U \sim 0.4$ for $i1=2$ or at  $U \sim 2.0$ for $i1=4$
in a way opposite to the case of F coupling.  
The transitions occur at similar values of $U$ in both the F and AF couplings. 
However, $J$-dependence is opposite between them.
The systems with $i1=1,3$, and $5$ do not exhibit phase transitions 
in the  calculated range of $ -0.5 <J<0$ and $0 <U< 8$.

\begin{figure}[tb]
\begin{center}
\includegraphics[width=7.0cm]{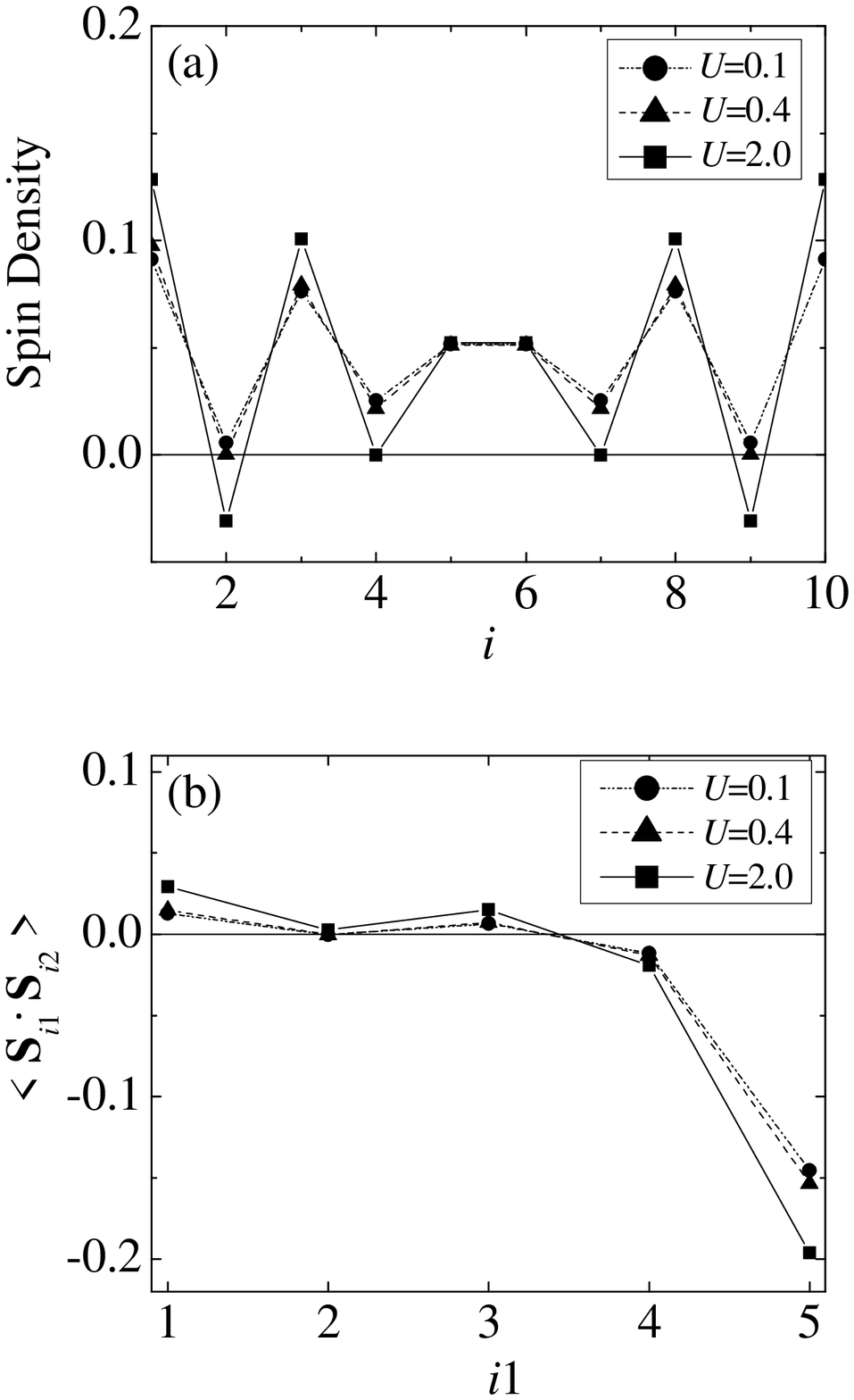}
\end{center}        
\caption{(a)Spin densities $\langle n_{i,\uparrow} - n_{i,\downarrow} \rangle /2$ 
         as a function of $i$; (b) spin correlations
        $ \langle \mathbf{S}_{i1} \cdot \mathbf{S}_{i2} \rangle $ 
        as a function of $i1$
        ($i2=N+1-i1$)
        for the doped state in the 1d Hubbard model  
        ($N=10, N_\uparrow=5, N_\downarrow=4$)
        with open boundary condition. 
        The dash-dot, dash, and solid lines correspond to $U=0.1,~0.4,~\text{and}~2.0$,
        respectively.}
\label{fig13}
\end{figure}

To understand the origin of these phase transitions, we calculated 
the doped states in 
the 1d Hubbard model  ($N=10, N_\uparrow=5, N_\downarrow=4$),
which is nothing but setting $J=0$ in our model. 
Their spin densities are plotted as functions of site($i$) in Fig. \ref{fig13}(a). 
The spin density is positive at every site
if the on-site Coulomb interaction is absent or quite weak. 
As the strength of  $U$ increases, 
spin densities become negative first at $i=2$ and $9$ for $U>0.4$,
 and then at site $i=4$ and $7$ for $U>2.0$.
The spin densities at the other sites
remain  positive irrespective of the strength of $U$. 
We also plot the spin correlation 
$ \langle \mathbf{S}_{i1} \cdot \mathbf{S}_{i2} \rangle $ 
with  $i2=N+1-i1$ in Fig. \ref{fig13}(b). 
The correlations are positive for $i1 = 1$ and 3, 
negative for $i1 = 4$ and 5, and almost zero for $i1 = 2$. 

\begin{figure}[tb]
\begin{center}
\includegraphics[width=7.0cm]{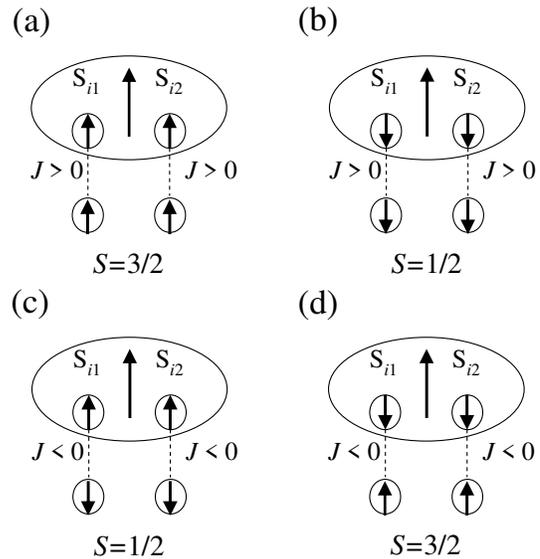}
\end{center}
\caption{Schematic picture of intramolecular spin alignment of doped system.
In the case of $J>0$:
(a) $S=3/2$  and (b) $S=1/2$;
in the case of $J<0$:
(c) $S=1/2$ and (d) $S=3/2$.}
\label{fig14}
\end{figure}

Returning to our original model with the localized spins attached,
we can understand the position-dependent alignment of the spins shown
in Figs. \ref{fig8} and \ref{fig9} from the spin densities of the $\pi$ electrons shown
in Fig. \ref{fig13}(a). 
In the case of F coupling, each localized spin tends to be in the parallel alignment
with the $\pi$-spin at its coupling site. 
If the coupling site has a positive spin density,
the localized spin is in the parallel alignment with the hole spin
leading to a spin quartet, as illustrated in Fig. \ref{fig14}(a). 
On the other hand,
the radical spin is  in the antiparallel alignment with the hole spin 
if the coupling site has a negative spin density, 
resulting in a spin doublet as schematically shown in Fig. \ref{fig14}(b). 
In addition, we can also explain the critical values of $U$ 
in the phase diagrams of Fig. \ref{fig11} 
in the limit of $J \rightarrow 0$:
The transition point is close to the value of $U$ 
where the spin density becomes negative in Fig. \ref{fig13}(a). 
The spin alignment in the case of  AF coupling 
can also be understood in a similar way as 
shown in Fig. \ref{fig14}(c) and (d).
This picture is consistent with the  spin correlation 
$\langle \mathbf{S}_{i1} \cdot \mathbf{S}_{i2} \rangle $,   
which is positive for large $U$ at $i1=1, 2$, and 3 as shown in Fig. \ref{fig13}(b).
As for  $i1 = 5$, the spin alignment for both 
F and AF couplings can be explained based on the spin densities, although the spin correlation
$ \langle \mathbf{S}_{\text{i1}} \cdot \mathbf{S}_{\text{i2}} \rangle $ is negative.

The absolute value of the spin density at $i1=4$ is very small, and
the parameter set is close to the phase boundary as shown in Figs. \ref{fig12}(b).
The sign of spin density at this site can be changed to positive
by a finite electron-lattice coupling.
As a result, spin doublet appears in the case of AF coupling, as schematically shown in Fig. \ref{fig14}(c).
The spin correlation
$ \langle \mathbf{S}_{4} \cdot \mathbf{S}_{7} \rangle $ is negative as shown in
Fig. \ref{fig13}(b), and takes control of the spin alignment in the case of F coupling.
This spin alignment leads to antiferromagnetic
correlation between the localized spins, and results in a spin doublet.
Such situation appears also at $i1=5$ for the case of
stronger coupling in our previous paper \cite{huai}.

\section{Summary}
\label{summary}
In summary, the control of spin alignment by charge doping 
is studied in the theoretical model of polyene-based molecular magnets.
By mean of the exact diagonalization technique,
we study the spin alignment, the total spins and gap energies 
in the doped  and the half-filled states.

Low-spin to high-spin transition is realized in such polyene-based molecular
magnets by one-hole doping into the $\pi$-conjugated moiety.
The doping-induced spin alignment depends on the molecular structure 
and the strengths of  the  interactions.
Alternating behavior of spin alignment is demonstrated in the hole-doped systems
with growth of the chain length.
Variation of spin-coupling positions leads to 
different alignment among the localized spins and the hole spin.
To understand behaviors in doped system, 
we investigate the  dependence of spin alignment on $J$ and $U$ in detail. 
The antiferromagnetic and ferromagnetic exchange interactions have different effect on doped molecules.
The overall behavior of  spin alignment can be understood 
in terms of spin density and correlation of $\pi$ electrons, 
in contrast to the $\pi$  topological rule for  the half-filled case where 
antiferromagnetic correlation in  $\pi$ electrons governs  spin alignment. 

We have shown that charge doping is an effective way to realize
controllable spin alignment in molecular magnets.  
As an important step towards molecular designing, it is vital to 
study molecular magnets based on more realistic $\pi$-moieties,
\textit{e.g.} fused carbon rings even with heteroatoms inside. 
The spin-alignment control in such $\pi$ molecular magnets will be reported elsewhere \cite{huai3}.

Authors are very grateful to Prof. A. Izuoka, Prof. Y. Teki, and Dr. T. Kawamoto for valuable discussions.
This work was partly supported by NEDO under the Nanotechnology Materials Program.

\end{document}